# Rethinking category-selectivity in human visual cortex

J. Brendan Ritchie[1]
Susan G. Wardle[1]
Maryam Vaziri-Pashkam[1,2]
Dwight J. Kravitz[3,4]
Chris I. Baker[1]

[1] Section on Learning and Plasticity (SLaP), Laboratory of Brain and Cognition, National Institute of Mental Health, Bethesda, MD, USA
[2] Department of Psychological and Brain Sciences, University of Delaware, Newark, DE, USA
[3] Department of Psychological and Brain Sciences, George Washington University, Washington, DC, USA
[4] Division of Brain and Cognitive Sciences, Social, Behavioral, and Economic Sciences Directorate, U.S. National Science Foundation, Arlington, VA, USA

Corresponding author:
Chris I. Baker (bakerchris@nih.gov)

**Abstract:**

A wealth of studies report evidence that occipitotemporal cortex tessellates into "category-selective" brain regions that are apparently specialized for representing ecologically important visual stimuli like faces, bodies, scenes, and tools. Here, we argue that while valuable insights have been gained through the lens of category-selectivity, a more complete view of visual function in occipitotemporal cortex requires centering the behavioral relevance of visual properties in real-world environments rather than stimulus category. Focusing on behavioral relevance challenges a simple mapping between stimulus and visual function in occipitotemporal cortex because the environmental properties relevant to a behavior are visually diverse and how a given property is represented is modulated by our goals. Grounding our thinking in behavioral relevance rather than category-selectivity raises a host of theoretical and empirical issues that we discuss while providing proposals for how existing tools can be harnessed in this light to better understand visual function in occipitotemporal cortex.

# 1. Introduction

How does visual cortex help us make sense of what we see? One popular answer is that it does so in part via a tessellation of occipitotemporal cortex (OTC) into regions specialized for representing complex stimulus categories, including faces, bodies, scenes, tools, words, and object forms more generally (**Box 1**). This topography of cortical activation, often referred to as *category-selectivity,* is one of the great success stories of visual neuroscience – indeed, it has been central to much of our own research, both past and present. Theoretically, it is fundamental to how the field thinks about clustering of visual function in OTC and dovetails with the assumption, common to computational models of vision, that categorization and the ability to label what we see is a major information-processing goal of the ventral visual stream [1,2]. Empirically, category-selectivity has proven to be highly consistent across individuals and robust to considerable variation in experimental design and stimulus manipulations. Further, many exciting current research directions in visual neuroscience focus on the developmental origins of category-selectivity in both humans and primates [3,4] and its computational basis as modeled with deep neural networks (DNNs) [5,6].

For all the progress the field has made from the study of category-selectivity, we believe it is worth asking whether this framework is still leading the field in the right direction (**Box 1**). In this *Perspective* we suggest a different theoretical and empirical course for the investigation of visual function in OTC that centers on the flexible use of visual information towards behavioral goals, rather than stimulus categories.

A commonly accepted explanation for why OTC exhibits category-selectivity is that region-defining stimuli are "ecologically important" to different forms of natural behavior that are well adapted to our environment[7–10]. For example, faces and bodies are important sources of social information for interacting with con- and allo-specifics; scenes include path and location information for navigation; and tools are graspable objects that we use in manipulating our environment. However, if the apparent selectivity for these stimuli is explained by the complex behaviors they facilitate, then we propose the topography of visual function in OTC may be better described as directly coalescing around these associated natural behaviors, rather than stimulus categories as such. In other words, the functional organization of OTC is optimized for representing properties of the visible environment relevant to planning natural behaviors– or more simply, their *behavioral relevance*[11,12]. As we show, explaining how the visual function of OTC is organized around behavioral relevance ultimately requires moving away from the framework of category-selectivity, even as we continue to build upon its hard-won insights.

We first elaborate on our characterization of behavioral relevance and the unique explanatory challenges that it poses (Section 2). Second, we show how the default framework of category-selectivity is in tension with prioritizing behavioral relevance in our understanding of visual function in OTC (Section 3). Third, we suggest that the empirical evidence for category-selectivity, while robust and reliable, has led much of the field towards a limiting view of visual function in OTC (Section 4). Given these considerations, we propose an alternative model for explaining how OTC codes behaviorally relevant visual properties, which also accommodates findings associated with category-selectivity (Section 5). Finally, we suggest practical ways that existing tools can be harnessed to study behavioral relevance more directly (Section 6).

## 2. Putting behavior first

In our view, a fuller understanding of visual function in OTC will only be achieved by adopting a more ethological framework that prioritizes behavior in our thinking. A foundational theme in the fields of both ethology and neuroscience is that studying a complex biological system requires asking not just what an organism is trying to do, but also *why*[13–16]. From an ethological perspective the "why" of visual processing is the facilitation of complex, adaptive, natural behavior in a dynamic environment. This perspective requires conceptualizing the functional organization of OTC in terms of its role in processing properties of the visible environment in the service of goal-directed behavior [17,18].

The *behavioral relevance* of any given visual property is determined by the interaction between what we see and what we are trying to do. Imagine going for a jog in a park and encountering a woman walking her dog (**Fig 1a**). In such a situation you may pursue a number of different goals, such as continuing with your run or stopping to pet the dog. For either action the visible environment offers a wealth of useful information. If you choose to jog past the woman, you might consider the distance to the tree, the wetness of the grass, and how fast she is walking. The gaze direction of the woman, the firmness of her grip on the leash, or the friendliness of the dog, may also be import sources of information. These very same visual features may also be important, but to different degrees, if you wish to pet the dog. In which case, the apparent friendliness of the dog, the gaze of the woman, and her grip on the leash may take on greater significance compared to how fast she is walking, the state of the grass, or the position of the tree line. This simple example highlights why behavioral relevance is not a fixed feature of the visual environment, but results from the interaction between the environment and our behavioral goals[11,12].

The interactive nature of behavioral relevance poses two important challenges to visual function in OTC [9,19]. The first is *visual diversity*: a wide range of properties in the visible environment can be relevant to the same behavioral goal (**Fig 1a**). For example, if you choose to jog around the woman or pet the dog, relevant visual information may come from material (wetness of the grass), spatial (distance to the trees), social (friendliness), relational (gaze direction), affordance (secureness of leash), or dynamic (speed of walking) properties of the environment. However, the specific list of behaviorally relevant visual properties might look quite different if our goal stays the same, but we are in a different visible environment (e.g., jogging on an indoor track or petting a dog at a shelter). Thus, neural coding in OTC must be able to cope with the *visual diversity* of behaviorally relevant properties we encounter in our environment when engaging in natural behavior.

The second is *goal-dependence*: although the properties of the visual environment may remain stable, our goals are not, and shifting them changes how we represent and evaluate what we see. In other words, although the same properties of a scene may be relevant to different goals, what will change is the significance of those properties. For example, whether a dog appears friendly is important whether your goal is to avoid or pet it (**Fig 1A**). In the former case, you may wish to know whether it will lunge as you pass by, and if it does so, will it be motivated by excitement or aggression. But the dog's friendliness may otherwise be of low importance to your goal. If you are stopping to pet it, its temperament is of central importance to how you approach. Thus, neural coding in OTC must be sensitive to this *goal dependence* and differentially process potentially behaviorally relevant sources of visual information.

If the visual function of OTC reflects the need to facilitate different types of natural, adaptive behaviors, then our focus should be to explain how it addresses these two challenges and bridges the gap between our goals and incoming visual information. In this respect, there is an important difference between the way in which behavior is commonly incorporated into the existing framework of category-selectivity and the course correction that we are proposing here in terms of behavioral relevance (**Fig. 1b**). While the category-selectivity framework initially stemmed from considerations of ecological importance, the focus of study is primarily on the stimuli (e.g. faces, scenes) and with associated natural behaviors (e.g. social interaction, navigation) following category-specific processing (e.g., recognition) of those stimuli. In contrast, our proposal involves prioritizing the behavioral relevance of visual properties, so that we consider goal-directed natural behavior first, and the visual information relevant to the specific behavior second. Thus, in the next two sections we show how the dominant framework of category-selectivity has both helped and hindered our understanding of how OTC meets the twin challenges of visual diversity and goal-dependence.

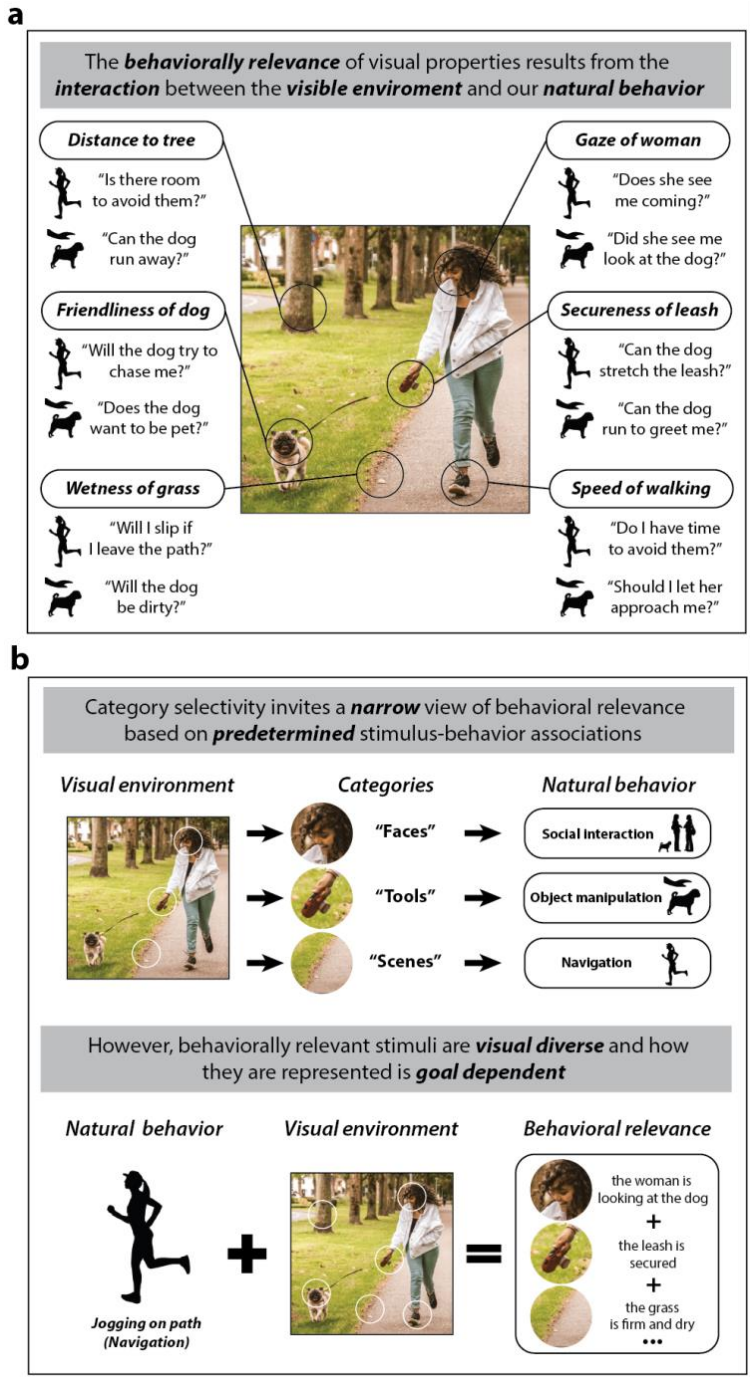

**Figure 1. Behavioral relevance as a framework for understanding visual function in occipitotemporal cortex.** (**a**) While we frequently experience digital renderings of visually complex natural images, our visual system is structured (through evolution and development) to encounter visual environments only during natural behavior. Many different aspects of the visual environment can be behaviorally relevant depending on the interaction between our behavioral goals and what we are looking at. (**b**) Within the traditional framework of category selectivity, the stimulus is parsed into visual categories (e.g. faces, tools, scenes) which are each associated with different broadly defined classes of behaviors (e.g. social interaction, object manipulation, visual navigation). In contrast, an ethologically inspired framework puts (classes of) natural behavior first (e.g. navigating by running around an obstacle), rather than the stimulus. Based on the behavioral goal, the relevant visual information is extracted from the stimulus— it is a product of both the visual stimulus and the natural behavior.

## 3. Studying category-selectivity is theoretically limiting

Recent work on category-selectivity has to some extent emphasized the importance of behavior to visual function in OTC. On the one hand, it has been proposed that OTC represents a variety of behaviorally relevant visual properties of region-defining stimuli beyond their category membership [7,10,20]. On the other hand, networks of purportedly category-selective brain regions in OTC have also been (re)described in terms of functional profiles directly related to classes of natural behavior. For example, tool-selectivity has been associated with a network of regions specialized for representing visual properties for object-directed action [21–23], and more recently face- and body-selectivity has been characterized as recruiting visual pathways specialized for social interaction [24].

These are important preliminary steps in changing how the field thinks about visual function in OTC. However, we believe further progress requires moving beyond the construct of category-selectivity (**Box 1**). As we argue in this section, category-selectivity: (i) invites us to view behavioral relevance through the lens of region-defining stimuli; and (ii) assumes categorization is a core aspect of information-processing in OTC. These two features of category-selectivity encourage a narrow view of visual function OTC that obscures the twin challenges of visual diversity and goal dependence posed by behavioral relevance.

### *3.1 The stimulus does not determine the behavior*

It has been proposed that how category-selective brain regions represent their preferred stimuli is shaped by the behaviors they facilitate [7,8,10]. In support of this, many studies have provided evidence that these regions represent the behaviorally relevant features of their region-defining stimuli. For example, so-called "scene-selective" regions have been shown to represent many scene properties highly relevant to navigation, such as the typical arrangement of objects in scenes [25], distance and openness [26,27], and navigational affordances [28,29], among other properties (for review, see [30,31]). In this way, category-selectivity emphasizes specific associations between specific region-defining stimuli (e.g., scenes) and natural behavior (e.g., navigation). However, this obscures the two most challenging features of behavioral relevance.

First, category-selectivity draws attention away from the visual diversity of behaviorally relevant properties in real-world visible environments. For example, when going for a run, we may deliberate about the best way to avoid an obstruction on the path, such as a woman walking her dog (**Fig. 1a**). In such a circumstance, the woman, her dog, and the dynamics of the leash are all highly relevant to the action we are taking, which is a form of navigation. Because category-selectivity specifically associates navigation with (non-agentive) scene properties, it limits consideration of navigationally relevant visual properties from other "categories" such as faces and manipulable objects because of their default associations with types of behavior such as social interaction and object manipulation (**Fig. 1b**). However, for natural behavior there is no simple mapping of visual stimuli to behavioral relevance as implied by category-selectivity.

Second, category-selectivity implies that behavioral relevance is a fixed feature of visual properties and not goal dependent. However, as we have emphasized, the very same visual properties can be important to different behaviors (**Figure 1**) [9,10]. For example, because faces, bodies, and animals all tend to be prelabeled as stimuli important to social interaction, they are typically ignored as sources of navigationally relevant information, as illustrated by the ubiquity of studies that use "scene" stimuli that are intentionally designed to exclude people and animals. Of course, some stimulus-behavior connections are immutable – we can only socially engage with other agentive living things, for example. However, the problem is that the flexible and goal-dependent nature of behavioral relevance is obscured by rigidly associating stimuli (e.g. agentless scenes) with a broadly defined behavior (e.g. navigation), as is the case with category-selectivity.

### *3.2 Categorization does not determine the behavior*

Constructing viewpoint invariant representations of object identity and category membership has often been characterized as a primary function of the ventral stream [1,6,32], with category-selectivity a byproduct of constructing such representations for ecologically important stimuli [2,33]. Recently it has been argued that, because OTC plays a central role in many behavioral tasks besides recognition, category-selective brain regions represent many properties that are distinct, and even orthogonal, to stimulus identity or category membership [7,10,20]. For example, head orientation and body configuration are important social cues but are incidental to determining individual identity [34,35]. Similarly, visual field position informs how we foveate or reach towards behaviorally relevant aspects of visual scenes, but retinotopic position is orthogonal to object category [36,37]. Under this broader view, categorization is just one of the functions implemented by OTC. However, this more expansive view still generates a distorted picture of the two main challenges posed by behavioral relevance.

First, categorization is sometimes considered a "task" of OTC comparable in status to forms of natural behavior like social interaction, navigation, and object manipulation [7,9,10]. In which case, it would delineate another type of goal dependence of visual function in OTC. However, putative forms of natural behavior, like navigation, are reflected in actions we plan and execute in the service of our goals (e.g., jogging along a path requires us to find a clear path and represent visual properties that suggest different navigational affordances). In contrast, visual categorization does not directly map to behavior in the real-world outside of experimental or basic educational settings any more than other forms of visual processing. However, when looking at an object like a dog, few would claim that representing something as oblong, brown, or furry are forms of natural behavior. Instead, in the context of natural behavior categorization is primarily a cognitive process that sometimes serves our goals and not a form of natural behavior in of itself [18].

Second, and in response to this problem, it is sometimes suggested that in order to plan behavior we have to first categorize what we are looking at [7,10]. In which case the visual diversity of behavioral relevance first depends on categorization as a filter. However, even if categorization helps us make sense of the visible world, it does not follow that categorization is required to determine which signals are behaviorally relevant [20,34,38,39]. For example, material properties are often important to behavior (e.g. whether the grass is slick or the leash is pliable in **Fig 1a**) and are represented in OTC [40–42]. But we can readily infer such properties directly when representing real-world environments without relying on categorization, even if material property estimation and categorization often interact [43]. Further, the information needed to inform behavior typically spans the putative region-defining categories (**Fig 1a**), making physiologically distinct representations of those categories inefficient. Thus, the diverse visual properties available during action planning need not be categorized in order to successfully inform behavior.

*3.3 Summary*

OTC represents the visible environment to inform natural behavior [18,44]. Given this, category-selectivity paints a misleading picture of the scope of behaviorally relevant properties that OTC represents as it places functionally undue importance on (i) region-defining stimuli; and (ii) visual categorization. These characteristics of category-selectivity obscure visual diversity and goal dependence, which are the main challenges facing visual processing in the service of behavioral relevance. Consequently, from an ethological perspective, it may be theoretically misleading to treat category-selectivity as central to the organization of visual function in OTC.

**4. Studying category-selectivity is empirically limiting**

By emphasizing the importance of behavioral relevance to visual function in OTC, it seems we are making a bold prediction: that there should be overlapping selectivity for a wide variety of stimuli, so long as they are relevant to similar natural behaviors. For example, regions characterized as face-selective should also respond to stimuli associated with other forms of category-selectivity, such as bodies, places, or tools, if they are relevant to social interaction. Yet, in apparent contradiction to this prediction, it is generally held that there is overwhelming evidence that these region-defining stimuli elicit dissociable loci of selectivity in OTC.

We do not dispute the robustness and reliability of category-selective effects per se (though see **Box 1**), which have also provided a key frame of reference for comparing the topography of OTC across participants, experimental designs, and research teams [45,46]. However, that robustness has led to a strong focus on finding category-selectivity and interpreting results within its context, leading to two key problems. First, many designs drastically under sample the diversity of possibly behaviorally relevant stimuli in favor of enabling simple contrasts to define and investigate brain regions. Second, even when results do show reliable responses that are more complex (e.g., graded rather than categorical selectivity), interpretations tend to overemphasize apparent categorical differences. Thus, the empirical base for category-selectivity has led to a restricted view of what drives clustering of visual function in OTC.

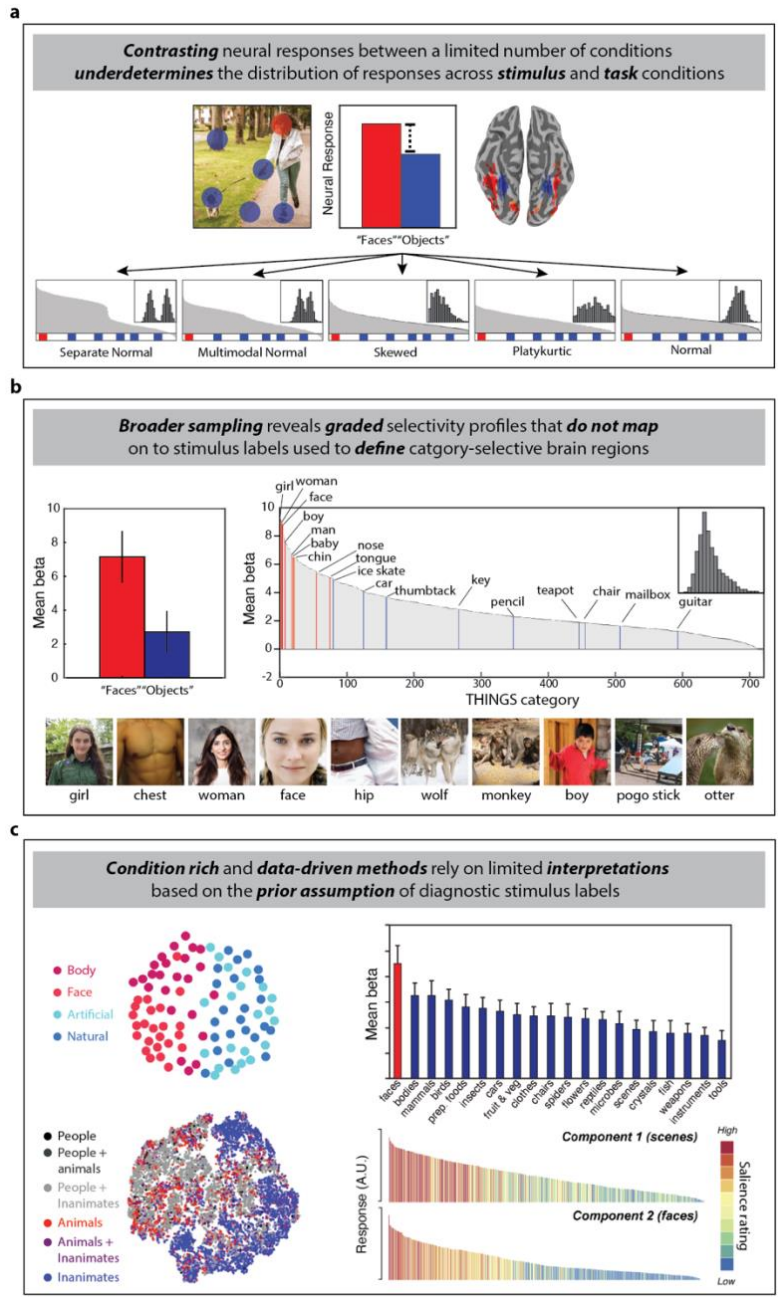

**Figure 2. How focusing on category-selectivity limits the investigation of behavioral relevance.** (**a**) Illustration of a common metric for measuring the visual selectivity of brain regions or neurons by comparing a small number of conditions based on their response magnitude between a target condition A/red and control condition(s) B/blue. The schematic charts illustrate hypothetical responses to a large number of stimuli ranked by magnitude, with the different underlying response distributions shown in the inset histograms (red and blue highlights on the x-axis indicate portions of the distribution that the target and control conditions might be sampled from). Only separate or multimodal distributions are indicative of a possible categorical division, which may also vary depending on behavioral goals. (**b**) Data from the THINGS fMRI dataset [47]. The response magnitude for 720 object concepts in the face-selective region of bilateral fusiform gyrus (n = 3). *Top-left*, a bar graph shows the mean beta averaged across voxels for the average of nine "face" concepts and nine "object" concepts. Error bars are +/- 1 standard deviation. *Top-right*, the distribution of responses (upper right) for all 720 concepts is shown ordered along the x-axis by response magnitude. The distribution is unimodal and skewed, rather than separate or bimodal. The location of the nine face and nine object concepts from are marked by red and blue vertical lines. *Bottom*, the top 10 of the 720 concepts that produced the strongest average response in the fusiform gyrus, and which are not obviously groupable by the label "face". (**c**) *Top-left*, 2D multi-dimensional scaling solution for the group-averaged patterns of neural responses to 92 stimuli, color coded by object type, in human ventral occipitotemporal cortex, which shows no bimodal clustering for the division between animate and inanimate objects [48]. *Top-right,* mean beta weights for 20 stimulus types in face-selective right fusiform gyrus [49]. Stimulus types are color coded based on whether they are faces (red) or not (blue). The responses suggest a skewed, unimodal distribution. *Bottom-left,* group-averaged t-SNE embedding of the neural responses for 10,000 images of complex scenes in anterior ventral occipitotemporal cortex, color coded based on (co)occurrence of object types [50]. Coding of the embedding points by stimulus type does not show separate categorical clusters. *Bottom-right,* group averaged response magnitudes for two neural components for 515 images of complex scenes, color coded based on the saliency of "scenes" and "faces" in the images [51]. Response profiles are unimodal and skewed, not separate or bimodal.

## *4.1 Limitations in experimental design*

Since studies of category-selectivity tend to selectively sample stimuli in their experimental designs, any results they report underdetermine the visual selectivity exhibited by OTC [52,53]. This is best illustrated by classic univariate analyses used to functionally localize category-selective regions of interest (ROI) in OTC. Typically, this involves comparing the magnitude of response between pairs of stimuli such as faces vs. objects, which results in a map of stimulus preference across ventral OTC — some cortical areas respond more to faces, and other areas to objects (**Fig. 2a**). This ubiquitous approach to defining functional ROIs presumes that these experimental conditions sample from different random distributions (e.g. separate normal distributions), however, when only two (or a few) conditions are considered, an apparent categorical difference between faces and objects could originate from many selectivity profiles in ventral OTC including continuous distributions (e.g., multimodal normal, skewed, platykurtic, normal) which have no categorical boundary (**Fig. 2a**). Thus, the apparent categorical response difference between faces and objects in ventral OTC may be an artifact of undersampling the underlying selectivity distribution. Simply put: one cannot infer selectivity from selective sampling – especially if we wish to characterize the diversity of visual properties in the environment relevant to natural behaviors.

Selective stimulus sampling is pervasive in the evidential base for category-selectivity, which includes a large body of work that has used a variety of methods to corroborate univariate findings. These include decoding the differences between patterns of neural responses for region-defining stimuli using machine learning classifiers [54,55], identifying differential patterns of functional and structural connectivity for region-defining stimuli [56,57], or selectively inhibiting neural responses for region-defining stimuli using transcranial magnetic stimulation (TMS) or microstimulation [58–

[61]. The same limitation even applies to clinical neuropsychology studies, which historically have been interpreted as providing some of the most compelling evidence for category-selectivity. For example, in both acquired and developmental prosopagnosia ("face blindness"), when behavioral performance is studied systematically in relation to other stimuli, face-specific deficits tend to be the exception rather than the rule [62,63].

Selective sampling is also a persistent feature of research on category-selectivity. This is best illustrated by two exciting research directions in visual neuroscience: the developmental basis of category-selectivity [3,4] and the modeling of category-selectivity with DNNs [5]. In the case of development, the same restricted set of stimuli, and comparisons, are still used whether they are studies of category-selectivity in infants and children [64–66], the congenitally blind [67,68], or non-human primates that have been reared with selective visual experience [69]. In the case of DNNs, several recent studies purport to show that topographic differences similar to those exhibited in OTC emerges in DNNs when trained to categorize region-defining stimuli [6,70,71]. However, despite the cutting-edge nature of these AI architectures, when it comes to the evidential base for category-selectivity these DNNs are employed to verify, rather than test, the same simple picture introduced by traditional univariate neuroimaging methods based on selective stimulus sampling. Because of this, they offer limited insight into the computational basis of clustering of visual function in OTC.

Again, we do not dispute the importance of these findings; indeed, it is only because of the empirical foundation that they provide that we have evidence that OTC codes for behavioral relevance. However, if we wish to understand how OTC might code for behaviorally relevant stimuli that are visually diverse, or how this coding is modulated by our behavioral goals, then the selective sampling that is paradigmatic of studies on category-selectivity is ill-suited to this explanatory endeavor.

*4.2 Limitations in interpretation*

The obvious solution to the selective sampling problem is to use larger, more diverse, stimulus sets. Going further, one could also try to minimize bias in the interpretation of results by adopting data-driven approaches to analysis. Indeed, studies that adopt these approaches purport to provide confirmation of category-selectivity in OTC. We agree wholeheartedly with the importance of both methodological shifts – although they do not directly address the challenge of goal dependence [72]. However, it is our contention that these studies do *not* provide clear confirmatory evidence of category-selectivity. Instead, the framework of category-selectivity has limited what analyses are performed and how findings are interpreted, even if results are more in line with predictions of the ethological framework we are proposing.

First, examining the response profile in putative category-selective brain areas for a much broader range of stimuli supports the idea that apparent categorical differences are frequently a result of undersampling. For example, when we plotted fMRI data from the THINGS dataset [47] in terms of the average response to nine "face" concepts, and nine "object" concepts in a face-selective region of the bilateral fusiform gyrus, we obtained the classic (apparently) categorical result of a much stronger response to faces over objects. (**Fig. 2b**, left panel). However, when we plotted the response to all 720 THINGS concepts separately and ordered them in terms of response magnitude, it revealed a skewed unimodal distribution with no discernable categorical boundary (**Fig. 2b**, right panel). Interestingly, the top 10 stimuli concepts with the strongest response magnitude include some faces, but also animals and body parts (**Fig. 2b**, bottom panel). These data are consistent with a higher response magnitude to stimuli relevant to social interaction (persons and other agents) in OTC, but not with a categorical difference in response to faces versus other objects [73,74].

Other studies using "condition rich" designs often select stimuli with the framework of category-selectivity in mind, resulting in a disproportionate number of exemplars from region-defining "categories" [48,75]. Even with this sampling bias, it is striking to note that the results of these studies do not clearly support categorical differences even in brain regions defined by selectivity to a particular stimulus, such as "faces" (**Fig 2c**). When the response magnitudes for multiple stimuli are plotted, the distribution is again typically graded, often without any clear category boundaries in the representational space or a step difference in response magnitude [49,75–77]. One concrete example of this is object animacy, which is thought of as a categorical property associated with face, body, and animal stimuli, but is represented in a graded and continuous fashion in regions of OTC [78–81]. Thus, even with broader stimulus sets, the interpretation of results is often influenced by the assumptions of the category-selectivity framework, even when they match the illustrative analysis in **Fig 2b**.

Second, data-driven methods have the potential to avoid methodological biases when they are used to analyze large neuroimaging datasets consisting of neural responses to thousands of visual stimuli that were not selected based on prior assumptions of category-selectivity. A striking example of this is provided by recent studies using the Natural Scenes Dataset (NSD), which consists of high-resolution fMRI responses and behavioral annotations for thousands of complex real-world scene images (**Fig 2c**)[50]. These studies found that the variation in the fMRI signal was explained by components or response profiles that seemed to match preexisting forms of category-selectivity OTC (e.g., for faces), and identified a less commonly recognized form of selectivity for food images [51,82,83] – though see [84]. These results have been interpreted as confirming the existence of category-selective brain regions by combining big-data with unsupervised analysis methods [85].

We believe the results of these NSD studies is consistent with a different interpretation, as illustrated by the supposedly "discovered" food-selectivity. Inspection of the graded response profiles reported by these studies show no categorical step [51] and all three studies localized food-selectivity both medial and lateral to face selectivity in the ventral OTC [51,82,83]. These expanses of cortex are also associated with clustering of selectivity for tools [22,23]. This similarity of clustering of selectivity for tools and food items makes sense from an ethological perspective, since in both cases we manipulate them with our hands in stereotyped ways during actions such as hammering a nail or stuffing our faces. In light of this, overlap in selectivity is to be expected, much as has been observed for hands and tools [86,87]. A recent study also provided direct evidence of this overlap for food- and tool-selectivity.[88] Thus, the NSD results are compatible with an alternative interpretation in terms of behavioral relevance rather than category-selectivity and demonstrate that even when using data-driven methods, prior assumptions of category-selectivity can influence how subsequent results are interpreted.

*4.3 Summary*
Studies of category-selectivity offer robust and replicable empirical findings, however the apparently categorical nature of response in OTC is likely a byproduct of undersampling the underlying distribution of selectivity for behavioral relevance. Recent approaches leveraging larger stimulus sets and data driven methods have the potential to avoid the issues inherent in selective sampling, as long as category-selectivity is not an implicit assumption in the experimental design or the interpretation of subsequent results. To understand how OTC supports natural behavior requires developing an alternative and more ethological framework, which we outline in the following sections.

**5. Charting behavioral relevance in OTC**

Despite our criticisms, there are several reasons the field has progressed using category-selectivity as a model of visual function in OTC (**Box 1**). First, it clearly delineates what kinds of complex visual properties are prioritized by OTC when representing real-world environments (e.g., faces and scenes). Second, it localizes the neural representation of these visual properties to networks of relatively discrete brain regions that encourages their isolation and probing through direct causal interventions. Finally, the topography of these purported brain regions is seemingly constrained by the functional organization of the visual system and the brain more generally [3,4]. If category-selectivity paints a theoretically and empirically limiting picture of how behavioral relevance drives visual function in OTC, as we have argued, then what is the ethologically inspired alternative *model* of visual function in OTC?

So far, we have suggested that the spatial clustering of neural responses typically described in terms of category-selectivity instead reflects the role of OTC in facilitating different forms of natural behavior. Furthermore, we have proposed that this clustering does not reflect coding for properties of region-defining stimuli but rather coding for behavioral relevance. In this section we build on this sketch by proposing a model of: (i) what behaviorally relevant visual properties are represented; (ii) how the representations of these properties are implemented in OTC; and (iii) which factors constrain the topography of these representations in OTC. In each case, we also suggest how to reinterpret the compelling picture offered by category-selectivity in light of these alternatives.

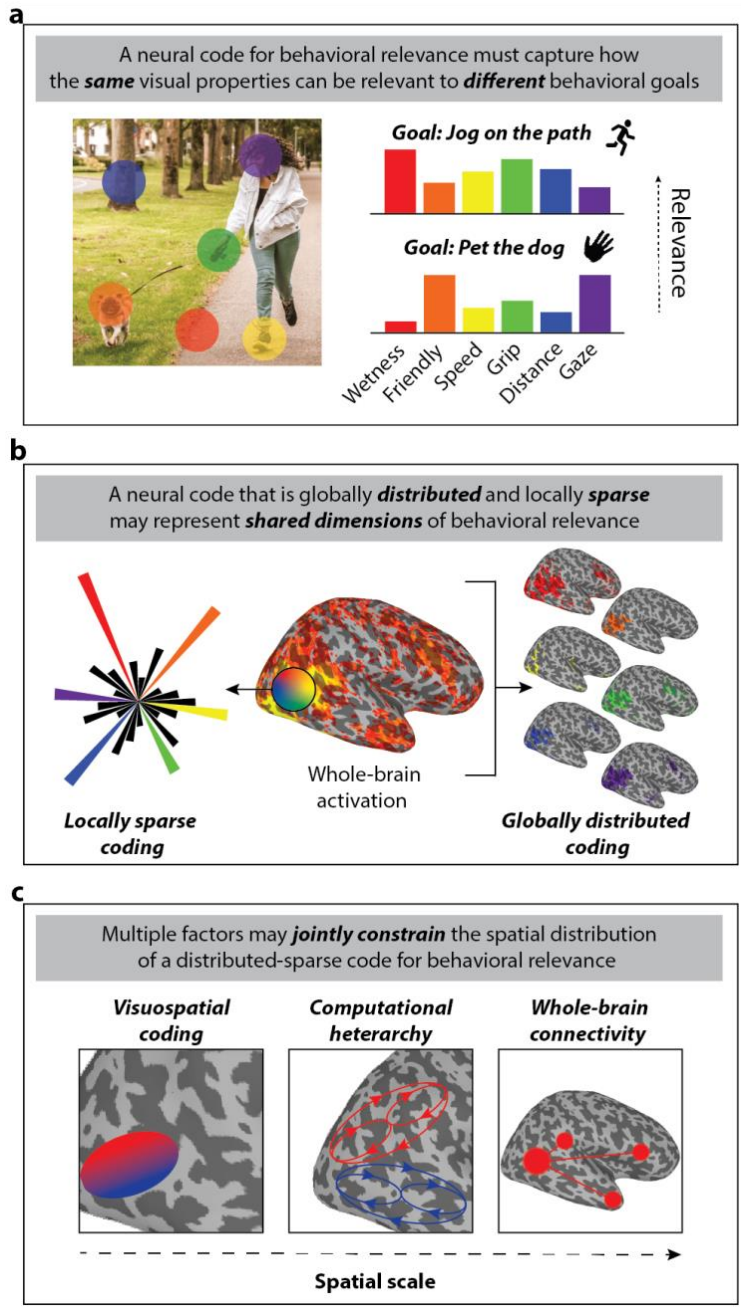

**Figure 3: A proposal for how visual cortex represents behavioral relevance.** (**a**) A heterogenous collection of visual signals can be collectively relevant to the same natural behavior (e.g., jogging on a path). These same signals may also be relevant to a very different behavioral goal in the same environmental context (e.g., petting a dog), but to different degrees. (**b**) Behaviorally relevant visual dimensions are likely to be represented in a widely distributed fashion throughout OTC and potentially visual cortex more generally, which is here depicted by color patches across the cortical surface showing where, hypothetically, the dimensions significantly explain some variance in the neural response. However, at any cortical locus, only some of these dimensions will be represented, which is depicted using a radial plot in which bars indicate the hypothetical degree of explained variance for different dimensions at cortical loci in OTC. In this respect the code is globally distributed but can be locally sparse. (**c**) Factors that determine cortical organization that may constrain the sparse coding of behaviorally relevant dimensions in OTC.

## 5.1 What visual properties are behaviorally relevant?

With respect to what is represented in OTC, the answer is simple: look to the behavior. We believe the forms of natural behavior associated with category-selective brain regions, such as social interaction, environment navigation, and object manipulation, are fitting starting points for inquiry [38,89]. But whatever natural behavior one starts with, whether jogging on a path or petting a dog (**Fig 3a**), the range of behaviorally relevant visual properties in the environment can only be identified in the course of specific actions.[90,91] Recognizing the active nature of vision is part of the impetus for emphasizing the two explanatory challenges posed by behavioral relevance that we have highlighted throughout (**Fig 3a**): visual diversity and goal dependence. These two features of behavioral relevance not only provide a critical grounding for how we should study visual function in OTC, they also make two crucial predictions: namely that there *is* graded, broad selectivity for a wide range of visual properties in portions of OTC previously associated with category-selectivity; and further, that the spatial extent and magnitude of this selectivity is modulated by task goals that recruit processes specialized for different natural behaviors.

How then should we think about the sorts of stimuli commonly associated with category-selectivity effects (e.g., faces and scenes)? As we have alluded to, these stimuli may be inherently important for certain forms of behavior, regardless of goal context [38]. For example, persons and animals are what we socially interact with in our environment, whatever our specific goals, or the present environment. Thus, labels like "face" and "body" point to stimuli of high relevance for (social) interaction, but as we have emphasized, the visual function of clustering in OTC is not to represent only these specific stimuli as such, but their behaviorally relevant properties and other parts of the visible environment. Furthermore, these same properties can be represented in different ways in the service of goals that fall within different behavioral domains. For example, gaze direction may be an important cue when we initiate social interaction, but it also constrains our navigational affordances when running on a crowded path (**Fig 3a**). Crucially, there is no sense in which a label like "face" or "body" reflects the default starting point for how we parse real world scenes for socially relevant cues. Instead, the "basic" or entry level at which we discern what visual properties are relevant to our goal is mutable, and influenced by context and experience, as is the case with basic levels in taxonomic hierarchies [92,93]. For example, a group of people interacting, or an agent performing a behavior, will produce many important visual properties relevant to our social goals (**Fig. 3a**). The full range of visible properties we observe in order to interpret the behavior of social agents are more indicative of the role of OTC in the larger domain-specific network for social cognition than whether they have a "face" or "body" or their category-membership [24,94].

*5.2 How are behaviorally relevant visual properties represented?*

If behaviorally relevant properties are visually diverse, and their relevance comes in degrees modulated by our behavioral goals, then how are they represented? Minimally, an appropriate neural code would require representations of sufficient dimensionality to represent the requisite diversity of visual signals, but in a way that can be reshaped to fit our differing behavioral goals. Here we adapt a model of visual function based on the analysis of the THINGS-fMRI data that originates in our lab [47]. Previous work has shown that cognitive representations of the similarity relations between the objects depicted in the THINGS images can be reduced to a small number of dimensions [95]. When mapped to the brain, these dimensions predict neural responses across large swaths of the cortical surface [96]. However, the cortical extent of this coding is not uniform across the dimensions, since at any given locus only some of the dimensions are represented (**Fig 3b**). In other words, there is heterogeneity due to varying levels of *sparse* coding across the cortical surface; i.e., the degree to which selectivity in a region is driven by only a few dimensions [96].

These results for the THINGS-fMRI data are similar in spirit to those of other studies that also suggest widely distributed, sparse coding across visual cortex, but for different dimensions related to simple visual features or semantic properties associated with stimulus categories [97,98]. However, we are proposing that the dimensions represented by this type of code are not derived from judged similarity, specific visual features, or semantic properties, but rather reflect degrees of freedom among many correlated visual properties in real-world scenes relevant to planning different forms of natural behavior (**Fig 3a**). For example, as mentioned earlier, evidence suggests "animacy" is not represented in visual cortex as a binary attribute of a stimulus, but a dimension that may encode several properties such as the agency or humanness of an object in graded fashion across the cortical surface [79,81,99]. Under our view, this dimension may be one of many that capture sources of variation relevant to different behavioral domains for which "animacy" provides a convenient, if potentially misleading, label [96]. We believe this type of sparse-distributed coding model has the potential to meet both of the challenges presented by coding for behaviorally relevant signals in OTC (**Fig 3b**).

First, this nascent model can explain how OTC codes for visually diverse properties. On the one hand, very different kinds of visual properties (e.g., various low, mid, and high-level visual properties) might be jointly coded along different dimensions of behavioral relevance. On the other hand, visually distinct objects (e.g., faces and bodies or hands and tools) show similar distributions of activity because they are high along the coded dimensions important to domains like social interaction and object manipulation. Second, this model can account for how behavioral relevance is modulated by our goals. A widely distributed representation allows for a behaviorally relevant dimension to be represented in portions of OTC that are part of many behavior-specific networks. Change in goal context will manifest as a shift in the distribution of the coding for the dimensions, as suggested by effects of task-based attention [94,100]. For example, we would expect the same sort of shift when visual signals normally associated with social interaction become relevant to navigation (e.g., where an agent is looking), this may shift the coding so that agent-related dimensions are now more strongly represented in regions of OTC that sparsely code for configuration information about scenes (**Fig. 3a**). Evidence supporting this prediction comes from studies that trained subjects to parse fonts made from face and scene images as opposed to more conventional orthography. Viewing the image fonts produced greater responses in regions of left ventral OTC already associated with reading words composed using more conventional orthographies [101,102].

How does the idea of distributed-sparse coding compare to the standard picture afforded by category-selectivity? On the one hand, according to the distributed-sparse coding model, purported category-selective regions can be described as clusters of relatively sparse coding for dimensions that are canonically relevant to particular classes of natural behavior [96] as opposed to stimulus categories and their properties [103]. For example, results obtained with the THINGS-fMRI data already show that sparse coding of similarity judgment dimensions recovers the locations of purported category-selective brains regions [96]. On the other hand, within the model any representation of these dimensions outside of sparse coding regions, like that suggested by the graded responses in (**Fig 2b-c**), are not considered aberrations of sampling noise but signals to be understood [104]. In this respect, the model recalls the earlier debate concerning whether selectivity for stimuli like faces and scenes is locally "modular" or widely "distributed" outside of category-selective regions [54,105]. In our view, there may be no tension between these alternatives since the proposed code is both widely distributed and locally sparse.

Our model is also similar in spirit to other characterizations of the organization of visual function in OTC. First, it has been hypothesized that there is graded selectivity across hemispheres that helps to explain specialization for different behavioral domains, such as the right vs left laterization for social interaction vs reading [106]. We conjecture that the same graded selectivity occurs within hemisphere as well, across multiple classes of natural behavior, in line with the premise that selectivity gradients are a ubiquitous organizing principle in the brain [44,107]. Second, there is evidence of extensive cross-talk between the dorsal and ventral visual streams depending on stimulus and task demands [108,109]. In this way, behaviorally relevant visual properties for reaching are represented by the ventral stream (e.g. complex shapes with more than one principal axis) and that information is passed on to the dorsal stream to guide action. We conjecture that the same sort of information sharing may occur between different clusters of sparse coding, which may help explain shifts in the distribution of response for different dimensions depending on stimulus and task conditions.

## *5.3 How are the representations constrained?*

Three factors have been proposed to constrain the locations of purported category-selective brain regions: maps of visual feature selectivity, the information-processing hierarchy of the visual system, and patterns of large-scale connectivity related to different behavioral domains (for reviews see [3,4]). For example, according to this multi-factor proposal, face-selectivity emerges during development in portions of feature maps that have foveal and convex curvature bias, are at a holistic representational stage of the processing hierarchy, and connect with social cognition regions elsewhere in the brain. Within our proposed model, so-called category-selective regions reflect clustering of sparse coding of behaviorally relevant dimensions. Furthermore, we have already pointed out that the evidence for these three factors, which comes from studies on the development of category-selectivity, faces the same selective sampling problem that plagues the evidential base for category selectivity (Section 4.1). Still, we conjecture that when characterized appropriately, each factor will likely play a role in explaining why broadly distributed but sparsely coded dimensions overlap to drive apparent clusters of selectivity.

First, patterns of (innately determined) connectivity between OTC and the rest of the brain provide the foundation for the emergence of whole-brain domain-specific networks, where "domain" refers to different types of natural behavior (or tasks) and the sort of information-processing required to carry them out — *not* a type of stimulus [110–112,89,113]. These networks have been identified for the different behavioral domains associated with forms of category-selectivity, including for navigation, social interaction, object manipulation, and reading, and are thought to constrain the location of putative category-selective brain regions in OTC [4,23,65,114–116]. However, unlike the framework of category-selectivity, we do not think that clustering of selectivity in OTC reflects stimulus-centered networks that interfaces with domain-specific ones; rather regions of OTC are the visual system component of these larger networks and represent visual signals relevant to the same behavioral domain, not the same stimulus. We would even go so far as to suggest that this picture is far more in line with the very idea of whole-brain domain-specific networks, which does not presuppose any form of stimulus specific selectivity[89]. Furthermore, the goal dependence of behavioral relevance implies that these networks also share information. For example, navigating a crowded room of people may recruit neural pathways that connect sparsely coded regions specialized for spatial processing to networks for social interaction.

Second, characterizations of the information-processing factor tend to presume a hierarchical characterization of the ventral stream focused on object and scene recognition [4,117], which is contrasted with the more action-oriented function of the dorsal stream [26,118]. Despite the ubiquity of this characterization, there are two ways it misconstrues the functional profile of OTC within the dual stream model of the visual system, which was also initially motivated by ethological considerations [17,119]. One is that a more encompassing characterization of the visual function of the ventral stream is that it represents a wide variety of stable visual properties of real-world environments in the service of cognitive processes like reward learning, memory, and decision-making to felicitate deliberation and also action planning [18,44,120]. The other is that the broad information-processing architecture of the ventral stream may be better described as being *heterarchical* [109,121,122]. For example, supposed low-level properties, like aspect-ratio, are represented in many regions of visual cortex associated with "high-level" properties like object category [123,124]. This alternative conception of the ventral stream may also be better suited to explaining sparse coding for a wide variety of potentially behaviorally relevant signals, which are represented in multiple ways depending on our goals (**Fig. 3a**).

Third, we believe visuospatial coding, in the form of overlapping visual feature maps (e.g. for spatial frequency, curvature, or retinotopy), likely provides a crucial scaffold for the representation of behaviorally relevant visual properties in real world environments [36]. What we dispute is that the visuospatial coding is tailored specifically to region-defining stimuli [3,103,125]. In fact, the evidence that visual features explain selectivity differences between category-selective regions is relatively weak [4,33,126]. Rather, it is plausible that visuospatial biases in sparsely coded regions are driven by statistical regularities related to how stimuli appear in the visual field (e.g., graspable objects in the lower periphery) during natural behavior, which are captured by different dimensions of behavioral relevance. The idea of visuospatial coding for behavioral relevance is compatible with early life experience with certain highly behaviorally relevant stimuli (e.g., the faces of caregivers) recruiting the protomaps of selectivity on which subsequent coding of behaviorally relevant dimensions is scaffolded [3,127]. Such scaffolding likely starts when infants are incapable of carrying out behavior, but are nonetheless learning a self-supervised model of real-world environments [128]. For example, the strongest evidence for the developmental role of visuospatial coding is that monkeys do not develop patches of face selectivity if they are deprived of exposure to faces early in life [3,69]. However, this lack of stimulus exposure also deprives them of typical face-directed social interaction and so is consistent with any feature-selectivity arising from learned associations between stimulus types and different behavioral domains [4,44].

*5.4 Summary*

Behavioral relevance encourages us to search for alternative models of visual function in OTC. We have proposed that OTC represents a diversity of visual properties based on their (goal modulated) behavioral relevance, and this may be implemented in distributed representations of behaviorally relevant dimensions with varying levels of sparsity in their topography, rather than circumscribed category-selective brain regions. Furthermore, functional organization in OTC will likely be constrained by the same factors (visuospatial coding, the ventral stream information-processing heterarchy, and whole-brain domain-specific connectivity) that have been posited to explain the locations of purported category-selective brain regions. We now turn to considering how this model can be explored empirically.

## 6. Behavioral relevance in practice

Given the above sketch of a coding model for behavioral relevance, how should we *study* visual function in OTC? Ultimately there must be a trade-off between ecological validity, practicality, and generalizability. Many methods, especially well-suited to studying how OTC codes for behavioral relevance, may require skills and resources that are only available to a few labs and so will only be implementable at scale through continued fostering of collaboration and Open Science [129]. However, taking a more ethological approach does not preclude the need for controlled experiments [130]. Nor does it require entirely new tools. The sort of naturalistic and data-driven methods that are increasingly being developed are a first, important step towards determining how behavioral relevance shapes visual function in OTC. But considerations of behavioral relevance still need to inform how we deploy these and other tools. In this final section we discuss several aspects of experimental design and how they should be shaped by considerations of the main features of behavioral relevance, its visual diversity and goal dependence. These aspects include: (i) stimulus design, (ii) tasks; (iii) participant selection; and (iv) data modeling and analysis. In each case, we highlight approaches that range in their feasibility.

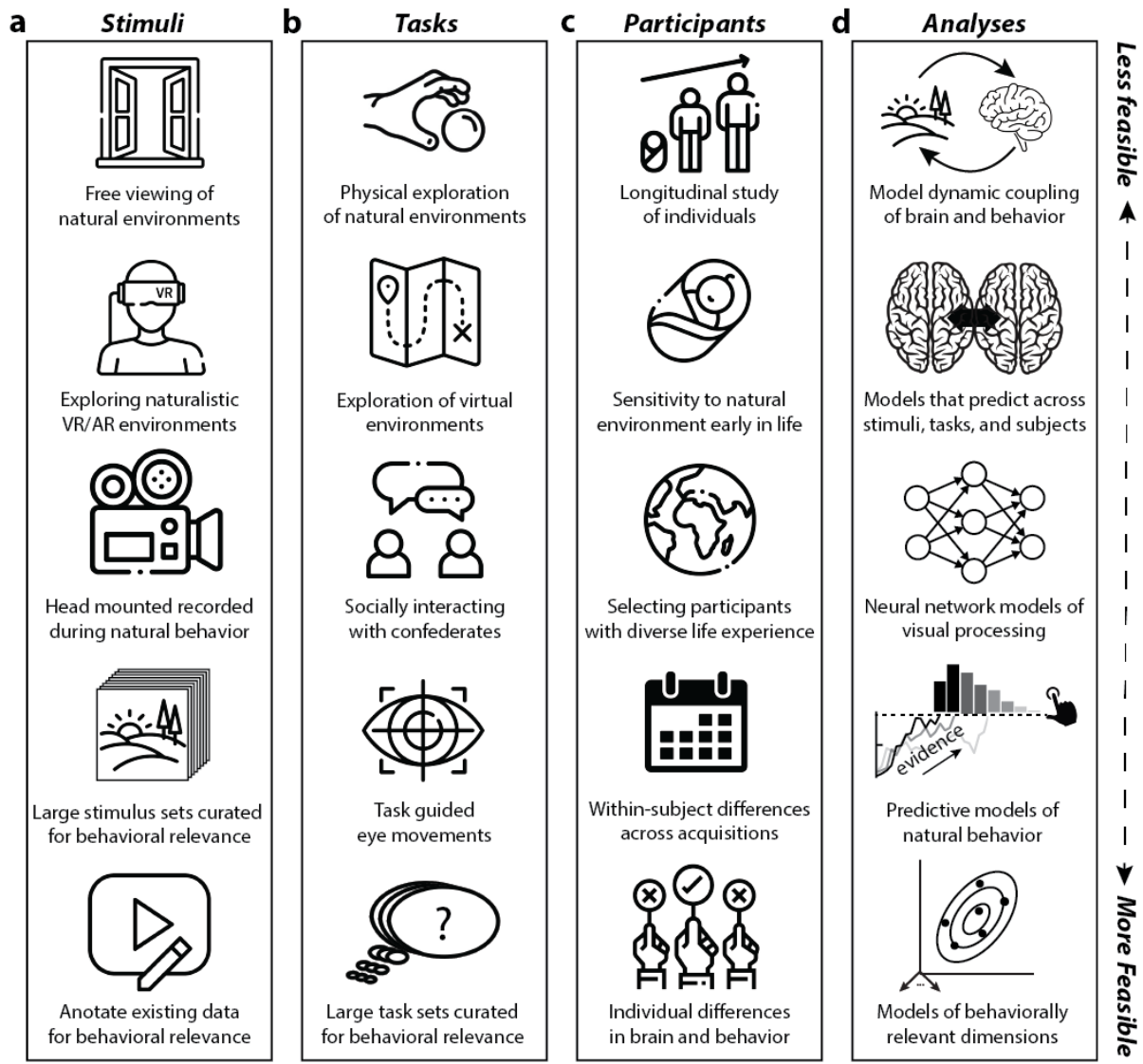

**Figure 4:** Suggested approaches for investigating the neural representation of behaviorally relevant signals in visual cortex. Behavioral relevance can be targeted through (**a**) stimulus design, (**b**) the nature of in-scanner tasks, (**c**) and participant selection by enhancing the naturalness of experimental conditions. However, different approaches will vary in the logistical challenges they present. While some practices are ideal, others are more practically feasible. (**d**) Similarly, the analysis and modeling methods vary in how ideal vs feasible they are for capturing sources of variation resulting from design choices illustrated in (**a**)-(**c**).

*6.1 Naturalistic stimuli*

Studying how OTC codes for behavioral relevance requires focusing on stimuli that are naturalistic, broadly sampled, and related to specific behavioral domains (**Fig. 4a**).

- Most ambitious would be mobile recording of dense brain and behavioral measures in real time in complex environments while participants perform natural behaviors [131,132]. While this approach potentially affords the ability to link brain measures with natural behavior, there are significant challenges due to experimental variability and the wide variety of factors that might influence neural activity.
- We can use active and immersive forms of stimulus presentation during acquisition including virtual and augmented reality to try and approximate natural viewing conditions during natural behavior [133–135].
- Continuous video recordings can be made while observers perform natural behaviors and used to develop stimulus sets as has been pioneered in infant studies [136,137].
- Large-scale short video or image stimulus sets of real-world environments can be curated to highlight behaviorally relevant environmental interactions such as social interactions and object manipulations [94,138–140] or to target non-categorical behaviorally relevant properties like affordances, materials, or object configurations [25,28,124,141].
- Existing large-scale datasets that use video or image stimuli, but were not designed to highlight behaviorally relevant visual properties, could be reannotated to try and target stimulus components related to different domains of behavior (e.g., not "is there a face" but, "Are people socializing in this image"?). [47,50,142–144]

## *6.2 Behavior during acquisition*

Since the behavioral relevance of visual properties is modified by our goals, it is critical that task manipulations during acquisition be the norm, rather than the exception, as is largely now the case in visual neuroscience [19,72]. To date, when task effects have been explored, the tasks that have been utilized are often behaviorally arbitrary (rather than naturalistic), limited in number, and focused on categorization or similarly constrained forms of judgment [145–148]. Generally, we should aspire to "task rich" designs that mirror the stimulus rich designs exemplified by recent large-scale datasets like NSD and THINGS, which will allow for systematic study of the effects of task focus and switching on neural coding in OTC [149].

- An ideal task manipulation would involve real-time recording while subjects engage in a wide-range of natural behaviors that may crosscut some intuitive divisions between behavioral domains (e.g., navigating through a room of people).[150]

- Although the forms of behavior possible during data acquisition is typically limited with human neuroimaging, it is possible to develop apparatuses that allow for overt behavior like reaching and grasping [134,151], or simulating environment exploration and navigation is possible with in-scanner virtual reality headsets [133].
- Live camera feedback during acquisition and "hyper-scanning" also allow for the possibility of volunteers to socially interact with a confederate in different scenarios [152,153].
- Eye movements have largely been treated as a source of noise when studying category-selectivity, but they are also modulated by stimulus and task demands, and can be used to obtain information about how we sequentially sample the environment during natural viewing [100,154,155].
- Even more common methods (e.g., button presses) can be used to target different behaviorally relevant aspects of complex natural stimuli, and provide the simplest approach for implementing task rich designs in which subjects alternate not between one or two, but potentially dozens of tasks tailored to the same or different behavioral domains [145–147,156].

## *6.3 Selecting participants*

The perceived behavioral relevance of stimuli varies across individuals, so it is critical to also consider individual and group differences in experimental designs. While variation in experience and knowledge will shape the neural response to behaviorally relevant signals, even when overt behavior seems similar, individuals may recruit different representations reflecting many possible solutions to the same task (**Fig. 4c**).

- Ideally longitudinal studies during development would allow for extensive comparisons within and between individuals as to how the visual system changes to accommodate increasing flexibility and variation in natural behavior [157].
- More achievable is to study behavioral relevance at different stages in development from infancy to adulthood, but without restricting stimuli to those associated with category-selectivity [64,66,158]. One option is to even re-annotate data acquired during continuous viewing in children for behavioral relevance [159].
- Specific groups of adults who vary in their personal experience can be targeted, as has been the case in work on visual expertise (e.g., chess players or bird watchers) [160–162]. The same group-level approach could be taken with respect to forms of experience relevant to natural behavior of interest and how this is impacted by cultural-specific knowledge [163] or personal familiarity [164,165].

- The topography of neural responses in OTC is assumed to be relatively stable within individuals. However, the phenomenon of "representational drift" [166,167] suggests that even seemingly stable behavioral responses may be underwritten by within subject changes in neural response over time, which may further be influenced by stimulus and task [104].
- In addition to selecting participants based on known differences in behaviorally relevant experience, it is crucial to study individual-level variation in the topography of neural responses within a single population. This has sometimes been investigated within the context of category-selectivity, but ideally requires behaviorally relevant stimuli and task selection [117,168,169], or direct links to natural behaviors like eye movements.[170]

## *6.4 Modeling and analysis*

Explaining how coding in OTC addresses the twin challenges of behavioral relevance (visual diversity and goal dependence) will also require us to change our perspective with respect to what sources of variation in neural data we are trying to explain. Many common and innovative models and analysis methods are already well-suited to this purpose (**Fig. 4d**).

- Since planning action is a *dynamic*, continuous process, the ultimate goal should be to model the ongoing coupling between the brain and environment during action without distinct stimulus or task events [171–173].
- We should strive for more ambitious *predictive* models derived from multiple datasets, across tasks, that allow for generalization to new contexts [174]. This would in principle entail developing models at a scale that captures large variation in stimulus conditions and tasks far beyond our current practices.
- Deep neural networks (DNNs) have provided an important tool for exploring hypotheses about visual processing and cortical specialization [5,70,71]. However, DNNs have often been benchmarked against datasets from category-selective brain regions [175,176], and so do not capture how OTC is specialized for behavioral relevance [7]. Examples include benchmarks such as Brain-Score (https://www.brain-score.org/) and the Algonauts Project (http://algonauts.csail.mit.edu/index.html). Going forward, any future models should be empirically checked against neural responses for datasets using richer stimulus and task designs (**Fig. 4a**), such as free viewing of infants [177]. Indeed, we believe that DNN models that are only trained on object or scene classification tasks will not do well on such comprehensive benchmark comparisons, and that new artificial neural network models will need to be developed that engage with the richer variety of behaviourally relevant signals that we have emphasized [178].

- We should not simply use behavioral models as separate predictors, but rather harness formal models to predict behavioral measures (e.g., reaction times, eye movements, and reach trajectories) from neural responses, as is common in so-called "model-based" approaches to neuroimaging [179,180].

- We can use existing approaches to capture the multidimensional nature of behavioral relevance. For example, encoding models have been applied to data collected during continuous viewing for large scale image sets [100,140,181,182] and representational similarity analysis allows for comparing second-order similarities in patterns of neural response to a wide variety of models, tasks, and participants [147,168]. However, predictors in these models should be based on behavioral relevance, not just familiar visual or categorical features.

*6.5 Summary*

As these suggestions make clear, developing a model of how OTC represents behaviorally relevant properties, and what neural processes delineate those properties, will require changes and innovations on multiple methodological fronts. This includes everything from stimulus design to the choice of tasks during data acquisition, through the selection of participants and the choice of models and analyses. Several recent advances are already moving the field in the right direction, and we are optimistic that the adoption of a more ethological framework for understanding visual function in OTC is well within our reach.

## 7. Conclusion

We have argued that recognizing the importance of behavioral relevance requires a shift in how we approach the study of visual function in OTC. In making our case, we have re-evaluated the theoretical and empirical foundations of the category-selectivity framework and taken lessons from the wealth of research it has produced. Though we have offered strong critiques and voiced misgivings (many of which also apply to our own work), we want to emphasize that the focus on category-selectivity has been incredibly successful, if mistaken – hence, we have characterized our discussion as a *rethinking* of the framework, and where it is taking us. After all, the correct inference to draw from our discussion is not that we "know nothing" about visual function in OTC.

Far from it. Rather, we believe the appropriate conclusion emphasizes not ignorance, but optimism: that we are close to an improved understanding of how the brain makes sense of what we see, and focusing more on how vision facilitates natural behavior offers us a route to get us there.

**Author contributions**

J.B.R and C.I.B conceived of the article which was further shaped through discussion with S.G.W., M.V-P. and D.J.K. J.B.R wrote the initial drafts, which were subsequently jointly revised and edited by all of the other authors.

**Competing interests**

The authors declare no competing interests.

**Funding**

This research was supported by the Intramural Research Program of the National Institute of Mental Health (ZIAMH002909 to C.I.B).

**Box 1: What is "category-selectivity"?**

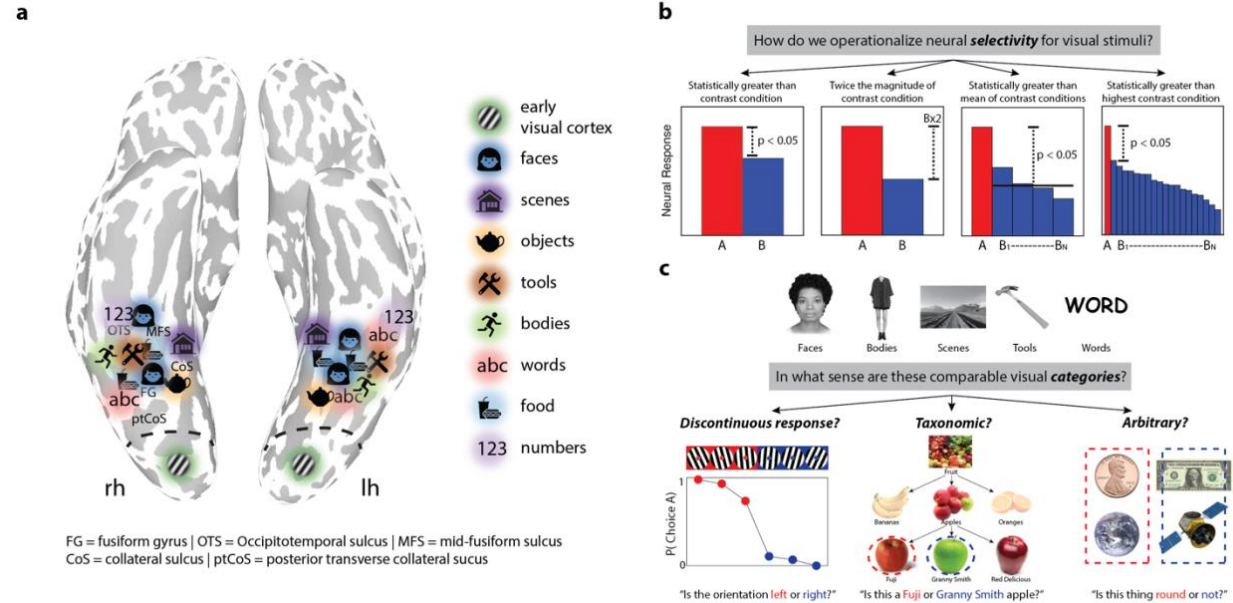

"Category-selectivity" is most strongly associated with clustering of neural responses in OTC for certain region-defining stimuli, as revealed with human neuroimaging (primarily fMRI). However, its origins are in neuropsychology and computer vision. For example, reports of face recognition impairment following lesion to the fusiform gyrus date back to the 19th century [183], and early work on semantic dementia described focal brain lesions that disrupted the ability to recognize or name specific types of objects (for a review see [184]). Similarly, the first primate electrophysiology studies of inferior temporal cortex found cells that responded most strongly to certain objects such as faces, bodies and hands [185–187]. Early human recordings from the cortical surface also revealed category specific responses, such as for faces [188,189] and letter strings [190]. These findings dovetailed with an interest, which remains strong, in visual recognition for categories within the computer vision community [13,191].

To date, "category-selective" responses have been reported for stimulus classes such as faces[192,193], objects[194,195], scenes[196], tools[197,198], bodies[199,200], words[201,202], numbers[203] and as recently proposed, food[51,82,83]. When the neural responses to these stimulus classes are contrasted with each other, they produce reliable peaks of activation in predictable locations that can be schematically mapped across the ventral surface of OTC (see the **figure, panel a**). While early reports described individual loci, subsequent work has revealed larger networks of regions, especially for faces [204], bodies [205], scenes [31], and tools [22], with additional peak responses on the lateral OTC surface (not shown).

Although such findings of "category-selectivity" are ubiquitous in visual neuroscience, its effects are not operationalized consistently and the construct often remains obscure despite some attempts at a clearer definition [33]. Here we briefly elaborate on both issues.

First, minimally a neural response or region of the brain is "selective" if it responds differentially and preferentially to a particular stimulus. Showing this usually involves comparing the neural responses between stimuli, but in practice many non-equivalent definitions are used that contrast the response to a target condition (figure, part **b:** A (red) = target; B (blue) = control condition(s)) [49,103,206]. However, the contrasted stimulus conditions are often selected because of historical precedent rather than a clear theoretical rationale. For example, much of the evidence for distinct face- and body-selective areas comes from studies that do not directly compare the responses between faces and bodies, even though they are part of the same behaviorally relevant object:

persons. In this respect, face- and body-selectivity may in fact reflect broader selectivity for persons, but this is obscured by standard procedures which fail to directly contrast them [74,207] (but see [208]). This type of concern is not new [52,53], but as we examine further in Section 4 (Fig. 2), it has important theoretical consequences for understanding visual function in OTC.

Second, the sense in which region-defining stimuli form comparable "categories" is rarely stated clearly. To make the issue more salient, consider that substantive notions of "category" do not readily group different region-defining stimuli together as comparable stimulus types (**figure, part c**). A category can arise from a *discontinuous* response to stimuli that vary continuously (e.g. left vs right orientation) [209,210]. Yet, stimuli like faces and scenes do not vary along visual feature dimensions such that they can easily be compared [103]. A category can also be seen as a component of a *taxonomic* system of knowledge (e.g. for fruit) [211,212]. However, there is no obvious taxonomic system that applies to both faces and scenes at different levels. Finally, categories can also be *arbitrary*, and so defined based on any rule or property that we can represent [212,213]. But this conflicts with the idea that OTC represents faces and scenes because they are ecologically important. Thus, it remains unclear what puts the "category" in "category-selectivity".